\documentstyle[12pt,epsfig,rotating]{article}

\begin{document}

\title{Initial \mbox{Friedmann} universe, its spatial flatness, matter creation,
and the cosmological term}
\author{N.~Tsitverblit\footnote{Address for correspondence: 
1 Yanosh Korchak Street, apt. 6, Netanya 42495, Israel;\newline
e-mail: tsitver@gmail.com} \\
School of Mechanical Engineering, Tel-Aviv University, Ramat-Aviv 69978, Israel}
\maketitle
\vspace*{-1cm}
\begin{abstract}
In the \mbox{Friedmann} equations, an infinite initial density
is avoided only when the universe is spatially flat. With such
equations being then valid when the scale factor $a=0$, the universe
must also be in the state of vacuum when $a$ is infinitesimal. Matter creation
thus largely comes from nonlinearity of the \mbox{Friedmann} equations
at a finite $a$. For the correspondence between such an initial vacuum 
and the \mbox{Friedmann} universe, therefore, a vacuum density is 
suggested to be gravitationally significant only when it could
also have a matter phase. The effective cosmological term is
then due to such a vacuum density. 
\end{abstract}
\section{\label{s:i}Introduction}
A major discrepancy between observed (presumable) consequences of the effective
cosmological term in the \mbox{Einstein} equations of general relativity and a
quantum-theory assessment of this term has been known as the (old) cosmological
constant problem. Different aspects of this issue have been repeatedly discussed
\cite{r:rp}. One recently emphasized additional element of such discussions is
also order-of-magnitude observational coincidence of the energy density
underlying the cosmological term with the matter energy density
\cite{r:rp,r:prs,r:trgl}.

Quantum mechanical understanding of vacuum contributions to the cosmological
term is a major fundamental issue \cite{r:rp,r:g,r:s,r:z}. Quantum theory is however
primarily applicable when a microscopic nature of matter is considered, as is statistical
mechanics in particular. Although gravity is also expected to be present in microscopic
processes, the most secured domain of application of the theory of general relativity
is described by phenomena of an intrinsically macroscopic nature. In this sense, this
theory is similar to the classical thermodynamics. Different aspects relating these two
theories have indeed been under intensive discussion \cite{r:tdm}. Respective theories
of microscopic and macroscopic matter description have to be consistent with each other.
At any energy scale, however, the universe as a whole is expected to remain infinite.
Its macroscopic description thus has to be permitted for any value of the scale factor.

If the \mbox{Einstein} equations of general relativity are correct on the universe
scale, therefore, the observations on the cosmological constant problems must have an 
interpretation in terms of these equations alone, for these observations may suggest 
their direct conflict with the \mbox{Einstein} theory. Such interpretations, however,
seem to be often based only on a phenomenologically (including observationally)
motivated specification of the spatial flatness and on additional phenomenological
assumptions giving rise to a negative pressure in the \mbox{Friedmann}
formulation for the \mbox{Einstein} equations. Apart from other effects \cite{r:rp},
in particular, these assumptions come from models of holography \cite{r:tdm}, vacuum 
decay \cite{r:vd}, bulk dissipation \cite{r:zmb}, or irreversible matter creation 
\cite{r:pn}. This note shows that a nonsingular and spatially flat universe with 
deflationary vacuum decay \cite{r:vd,r:gd,r:bp,r:ir} and an effective cosmological
term generally arises from the \mbox{Friedmann} form of the \mbox{Einstein} equations 
without additional assumptions.
\section{\label{s:iusf}Initial universe and its spatial flatness}
\subsection{\label{s:mo}Mathematical observations}
For the \mbox{Einstein} equations \cite{r:ll} with the
\mbox{Friedmann}-\mbox{Robertson}-\mbox{Walker} (FRW) metric \cite{r:kt}, let us consider
the acceleration and \mbox{Friedmann} equations (hereafter both referred to as the 
\mbox{Friedmann} equations) \cite{r:rp,r:kt,r:fn} (in the units with
the speed of light $c=1$), 
\begin{equation}
\frac{\ddot{a}}{a}=-\frac{4\pi G}{3}(\rho+3p),
\label{eq:f1}
\end{equation}
\begin{equation}
H^{2}=\frac{8\pi G\rho}{3}-\frac{k}{a^{2}}.
\label{eq:f2}
\end{equation}
Here $a$ is the scale factor, $H={\dot{a}}/a$ is the \mbox{Hubble} parameter, 
$G$ is the gravitation constant, $k=0,\pm 1$ is the space-curvature index, $\rho$
and $p$ stand for the overall energy density and pressure, respectively, and the
dots denote the derivatives with respect to time, $t$.

Let $\rho$ and $p$ be viewed as the functions of $a$.
Eqs. (\ref{eq:f1}) and (\ref{eq:f2}) are thus invariant with respect 
to both a shift in time (transformation $t\mapsto t+\tau$ for any $\tau$) and the 
direction of time (transformation $\tau+t\mapsto\tau-t$ for any $\tau$). Since $a=0$
is a minimal value of $a$ when the latter is considered as a function of either the
positive or the negative time, the existence of $\dot{a}_{|a(\tau_{0})=0}$ (regardless
of how the time moment, $\tau_{0}$, at which $a=0$ is defined) then requires that 
$\dot{a}_{|a(\tau_{0})=0}=\dot{a}_{|t=\tau_{0}}=\dot{a}_{|t=\tau_{0}-0}=
[da(\tau_{0}-t)/d(\tau_{0}-t)]_{|t=0}=-[da(\tau_{0}+t)/d(\tau_{0}+t)]_{|t=0}=
-\dot{a}_{|t=\tau_{0}}=0$. Eq. (\ref{eq:f2}) thus implies that the
density at $a(\tau_{0})=0$ could be finite \cite{r:gd,r:bp,r:ir,r:mr}
only if the universe is spatially flat, consistently with the observations
\cite{r:rp,r:prs,r:trgl}. In particular, application of the \mbox{l'Hospital} rule
for $a\rightarrow 0$ to Eq. (\ref{eq:f2}), along with Eq. (\ref{eq:f1}), then
suggests [allowing for $d()/da=\dot{()}/\dot{a}$] that the only equation 
of state permitted for $a\rightarrow 0$ is that of vacuum:
\begin{equation}
p_{v}^{0}=-\rho_{v}^{0},
\label{eq:v0}
\end{equation}
where $\rho_{v}^{0}$ and $p_{v}^{0}$ are the energy density and pressure when
$a\rightarrow 0$, respectively. [Eq. (\ref{eq:v0}) would not be affected if
Eqs. (\ref{eq:f1}) and (\ref{eq:f2}) also had a cosmological constant term
that is not a part of $\rho$ and $p$. That such a term has to be irrelevant
will be seen from below.] Along with Eq. (\ref{eq:v0}), Eq. (\ref{eq:f1}) 
thus implies that $\ddot{a}>0$ at $a=0$, i.e. that $a=0$ 
is a minimum of $a(t)$.

After a medium with the equation of state described by Eq. (\ref{eq:v0})
was first introduced in \cite{r:g}, such a medium has also been hypothesized in
\cite{r:gd,r:bp,r:ir} to describe an initial state of the universe that
does not meet the conditions for existence of a cosmological singularity
\cite{r:hpg}. Analogous possibilities for preventing the singularity also
predicted in \cite{r:hpg} to arise otherwise in the context of a black 
hole gravitational collapse have been considered in \cite{r:dm}.

For a finite $\rho(0)$ and Eqs. (\ref{eq:f1}) and (\ref{eq:f2}) being valid at
$a=0$, one could expect $[{d\rho/da}]_{|a\rightarrow 0}=O(a)$ to hold as well. Let 
us however also note that Eq. (\ref{eq:f2}) formally requires that $\rho(a)=\rho(-a)$.
Eq. (\ref{eq:f1}) then implies the validity of $p(a)=p(-a)$ as well. As for
$\dot{a}(\tau_{0})$ above, this means that 
\begin{equation}
[\frac{d^{(2n+1)}\rho(a)}{da^{(2n+1)}}]_{|a=0}=
[\frac{d^{(2n+1)}p(a)}{da^{(2n+1)}}]_{|a=0}=0,\,\,\,\,\,\,n=0,1,2,3,...
\label{eq:doz}
\end{equation}
have to hold. For $n=0$, Eqs. (\ref{eq:doz}) thus suggest that $\rho(a)$ and $p(a)$
are actually expected to have their respective extrema at $a=0$. Eq. (\ref{eq:v0})
could therefore also be obtained if one considers the $a\rightarrow 0$
limit of the energy-conservation equation,
\begin{equation}
\frac{d\rho(a)}{da}=-3\frac{\rho(a)+p(a)}{a}.\,\,\,\,\,\,\,\,\,
\label{eq:f3}
\end{equation}
This equation is implied by Eqs. (\ref{eq:f1}) and (\ref{eq:f2}) as well.

By differentiation of Eq. (\ref{eq:f3}) and using Eq. (\ref{eq:f3}) itself,
one could then obtain:
\begin{equation}
\frac{d^{2}\rho(a)}{da^{2}}=
-\frac{3}{a}\{\frac{d}{da}[\rho(a)+p(a)]+
\frac{1}{3}\frac{d\rho(a)}{da}\}.
\label{eq:d2}
\end{equation}
With the use of the \mbox{l'Hospital} rule in Eq. (\ref{eq:d2}) for $a\rightarrow 0$
as well as of the Eqs. (\ref{eq:doz}) for $n=0$, 
\begin{equation}
[\frac{d^{2}p(a)}{d a^{2}}]_{|a\rightarrow 0}=
-\frac{5}{3}[\frac{d^{2}\rho(a)}{d a^{2}}]_{|a\rightarrow 0}\,\,\,\,\,\,
\label{eq:d2z}
\end{equation}
is thus derived. Since $p(a)>-\rho(a)$ for $a>0$, therefore, Eqs. (\ref{eq:v0}),
(\ref{eq:doz}) for $n=0$, (\ref{eq:f3}), and (\ref{eq:d2z}) suggest that $a=0$ has
to be a maximum of $\rho(a)$ \{i.e., $[d^{2}\rho(a)/da^{2}]_{|a\rightarrow 0}<0$,
as implied by Eq. (\ref{eq:f3}) for small $|a|\geq 0$\} and a minimum of $p(a)$ 
\{i.e., $[d^{2}p(a)/da^{2}]_{|a\rightarrow 0}>0$\}.

When Eqs. (\ref{eq:f1}) and (\ref{eq:f2}) are valid at $a\rightarrow 0$,
their linearization with respect to $a$ at $a=0$ is legitimate, say upon
multiplication of Eq. (\ref{eq:f2}) by $a^{2}$ and of Eq. (\ref{eq:f1}) by
$a$. With $[{d\rho/da}]_{|a=0}=0$, such linearized Eq. (\ref{eq:f2}) and 
Eq. (\ref{eq:f1}) itself at $a\rightarrow 0$ lead to the equation of
state described by Eq. (\ref{eq:v0}) being approximately valid for an
infinitesimal $a$ ($\equiv a_{i}$ say) as well. In particular, the 
right hand side of the Eq. (\ref{eq:f3}) multiplied by $a$ differs 
from zero by a magnitude at least of the order of $a^{2}$, 
since $d\rho(a)/da=[d^{2}\rho(0)/da^{2}]a+o(a)$ for
$a\rightarrow 0$.
\subsection{\label{s:pi}Physical implications}
With a finite initial density when the universe
is spatially flat, the (adiabatic) thermodynamics incorporated in the
\mbox{Friedmann} equations thus actually requires that any (relativistic and 
nonrelativistic) matter significantly manifested for a finite $a>0$ macroscopically
behave for $a\rightarrow 0$ in compliance with the vacuum equation of state. Matter
creation could therefore be interpreted as a matter self-identification when
$a$ increases from $0$. This process has to become particularly intensive
when $a$ reaches such a (finite) value as renders the linearized
\mbox{Friedmann} equations inadequate and necessitates a 
manifestation of the nonlinear effects.

Since the \mbox{Friedmann} equations do not contain any quantum physics, 
which has been invoked for interpretations in \cite{r:gd,r:t}, the matter-creation
process is describable in classical terms alone. Based on the energy conservation
underlying Eq. (\ref{eq:f3}), such a possibility has generally been emphasized in
\cite{r:gd,r:bp}. Various phenomenologically constructed models of vacuum decay
into matter have thus been considered \cite{r:vd,r:bp,r:ir}. Alternative classical
interpretations of matter creation and acceleration have been based on models
of other different aspects of thermodynamic irreversibility 
\cite{r:tdm,r:zmb,r:pn}.

The physics of matter creation suggested
in Sec. \ref{s:mo} above might however also be viewed as
classical manifestation of a quantum phenomenon some of whose
aspects have been discussed in \cite{r:psz}. Such a phenomenon 
could be associated with production of permanent (real) matter particles
out of temporarily existing (virtual) ones when annihilation of the latter
is prevented by the effect of space expansion.
The (quantum) temporal violation of energy conservation by the time-energy 
uncertainty is then turned into a classical decrease of the energy density.
This decrease could be offset by growth of the overall vacuum energy due to
the expansion work. Such work is done against the vacuum pressure by the 
repulsive vacuum gravitation while it acts on the created matter.

It is interesting to note in this context that the two exponential
terms arising from (either of) Eqs. (\ref{eq:f1}) and (\ref{eq:f2}) 
and the (nearly) vacuum initial equation of state [approximately described
by Eq. (\ref{eq:v0})] at an infinitesimal $a$ might also actually correspond to
two directions of time \cite{r:s,r:skhp}, with respect to which Eqs. (\ref{eq:f1})
and (\ref{eq:f2}) are symmetric. Indeed, the term with the plus sign in the 
exponent has to vanish if $a$ is required to be zero when $-t=-\infty$.
In this case, $a$ grows with $-t$ being increased due to the term with 
the minus sign in the exponent. This latter term then has to vanish if
$a$ is required to be zero when $t=-\infty$. Growth of $a$ with $t$
being increased is then due to the term with the plus sign in the 
exponent. With such an interpretation, virtual particle-antiparticle
pairs at an infinitesimal $a$ are separated by the two branches of 
the CPT symmetry to eventually give rise to two different universes
dominated by real matter (where entropy increases with $t$) and 
real antimatter (where entropy increases with $-t$),
respectively \cite{r:s,r:skhp,r:aqn}.

A matterless vacuum with any homogeneous energy density could thus correspond
to a macroscopic manifestation of the zero-$a$ limit of a \mbox{Friedmann}
universe with the respective matter-density distribution, $\rho_{m}(a)=\rho(a)$,
for a large enough $a$. Such a $\rho_{m}(a)$ is expected to be the larger the
higher its vacuum density is at $a\rightarrow 0$. From this perspective,
matter does not have to disappear in the respective vacuum. The behavior 
of its particles is however expected to become such as mimicks
the behavior of virtual particles in vacuum \cite{r:gd}. 

With the above correspondence, suggested by the general relativity
theory, a growing $a>0$ stands for both the transition from vacuum 
to matter and the gravitational action of the residual vacuum and the 
created matter. In this context, a permanent part of vacuum density,
if any, could then have only a(n equal) constant space-time response, 
for such a part corresponds to $a=0$. A vacuum density thus has to 
be gravitationally significant only when it could also have a
respective matter phase. For the \mbox{Einstein} equations, 
this is discussed in Sec. \ref{s:cc} as well.

There thus seems to be no reason to {\it a priori} expect what is often referred
to as ''breakdown'' of the theory of general relativity when $a\rightarrow 0$. As
discussed above, the universe remains a(n infinite) macroscopic object that is capable 
of maintaining a finite density in this limit as well. Since the universe has to be
spatially flat to avoid an infinite initial density, the so-called ''flatness problem''
\cite{r:grn} does not have to exist either. In particular, this could question some
of the cosmological motivation for introduction of an inflation paradigm into the
early universe \cite{r:kt,r:inf}. Such a paradigm also seems to be inconsistent
with the physics of matter creation outlined above. The latter process starts
as soon as $a>0$, when $[d\rho(a)/da]a<0$ in Eq. (\ref{eq:f3}). Anticipated
difficulties for reconciling the inflation paradigm with such theories
as general relativity and thermodynamics have already been previously
emphasized \cite{r:infdg,r:infps}.
\section{\label{s:mc}Matter creation}
\subsection{\label{s:gc} General considerations}
The observed accelerated expansion \cite{r:rp,r:prs,r:trgl} is often
associated with the presence of a new form of energy density other than
a matter density in the universe. According to the standard ($\Lambda$CDM)
scenario \cite{r:rp,r:grn02}, such a hypothetic density also has to be
especially dominant for $a\rightarrow\infty$. In particular, Eq. (\ref{eq:f2})
suggests that the overall density of the universe could be greater than zero at 
$a\rightarrow\infty$ only when $\dot{a}_{|a\rightarrow\infty}\rightarrow\infty$ as
well, consistently with the standard scenario. (The velocity of space expansion is not
limited by the speed of light.) Using the \mbox{l'Hospital} rule in Eq. (\ref{eq:f2})
with $\dot{a}_{|a\rightarrow\infty}\rightarrow\infty$ and Eq. (\ref{eq:f1}) 
when $a\rightarrow\infty$, one then obtains:
\begin{equation}
p_{v}^{\infty}=-\rho_{v}^{\infty},
\label{eq:vinfty}
\end{equation}
where $\rho_{v}^{\infty}$ and $p_{v}^{\infty}$ are, respectively, the vacuum energy
density and pressure at $a\rightarrow\infty$.

In the absence of phase transitions between vacuum and matter, the energy
density satisfying Eq. (\ref{eq:vinfty}) still has to be independent of the scale
factor, including $a\rightarrow\infty$. This follows from Eq. (\ref{eq:f3}) for such 
a density component and from more general thermodynamic considerations \cite{r:ls}.
Had such phase transitions ended before the stage of accelerated expansion, therefore,
the observed acceleration would have been driven by the vacuum density identified by 
Eq. (\ref{eq:vinfty}). Such a scenario, however, implies the existence of a permanent 
part of the vacuum density that affects the universe dynamics without being the origin
of matter. This would be inconsistent with gravitational significance of only such a 
vacuum density as could also have a respective matter phase. Such inconsistency is 
due to the correspondence between a homogeneous matterless vacuum and the respective
\mbox{Friedmann} universe, as discussed in Sec. \ref{s:pi} above. In the framework of
this correspondence, therefore, vacuum decay has to be active at the acceleration stage, 
as in some phenomenological models of this process and interpretations of cosmological 
data \cite{r:vd,r:vdo}. Even if the rate of vacuum decay into matter is such as a 
gravitationally significant vacuum density would remain at any finite $a$, the
argument just used still implies $\rho_{v}^{\infty}=p_{v}^{\infty}=0$.
The zero overall density and pressure at $a\rightarrow\infty$ 
could not thus be avoided.

In the framework of Eqs. (\ref{eq:f1}) and (\ref{eq:f2}), emergence of 
the main part of the overall amount of current matter out of $\rho_{v}^{0}$
is possible only when the scale factor is larger than a finite value, $a_{0}$.
With such nonlinear matter creation being most intensive initially, a stage of
dominance of matter attractive gravitation \cite{r:rp,r:prs,r:trgl} could 
thus exist. To overcome this stage, the matter created by such a nonlinearity
then has to emerge with a sufficiently high initial velocity of expansion.
The corresponding initial value problem for subsequent evolution of the
universe could thus become similar to the boundary value problem with
a given overall (current) amount of matter in the standard cosmology 
\cite{r:rp,r:grn02}. For the cosmological scenario considered herein,
however, the high initial velocity of expansion acquires
a simple and natural explanation.

As the scale factor, $a$ has to remain physically meaningful even
when it is infinitesimal, i.e. say when $a=a_{i}$, and an equation of state closely
approximating Eq. (\ref{eq:v0}) is macroscopically maintained in the universe. Matter
could then be viewed as mainly test particles having only a higher-order infinitesimal 
effect on the equation of state, as seen from Eq. (\ref{eq:f3}). Such an effect
is however sufficient to allow a further growth of $a$ and then a substantial 
manifestation of the phase transitions significantly changing the equation
of state when the scale-factor magnitude triggering nonlinearity 
of the \mbox{Friedmann} equations, $a=a_{0}$, is reached.

For the reasons discussed above,
the existence of a gravitationally significant part of the initial
vacuum density is prevented so long as such a part is not the origin 
of any matter. Since $\rho_{v}^{0}$ at $a=0$ thus results from the 
zero-$a$ limit of all the matter that would eventually be
created (at an $a\gg a_{0}$), one could write:
\begin{equation}
\rho_{v}^{0}=\rho_{vr}^{0}+\rho_{vn}^{0}.
\label{eq:den0}
\end{equation}
Here $\rho_{vr}^{0}$ and $\rho_{vn}^{0}$ are the constituents of $\rho_{v}^{0}$
that give rise to what is manifested at $a>0$ (particularly for $a\geq a_{0}$) as
radiation and nonrelativistic matter, respectively. The symbols in Eq. (\ref{eq:den0})
thus represent the values of respective vacuum densities $\rho_{vr}(a)$,
$\rho_{vn}(a)$, and $\rho_{v}(a)=\rho_{vr}(a)+\rho_{vn}(a)$ at $a=0$. Such a
separation between $\rho_{vr}(a)$ and $\rho_{vn}(a)$ in the vacuum state at
$a=a_{v}$ (including $a_{v}=0$), however, does not imply the absence of
interaction between the respective components of matter after either
one of them has or they both have been created at 
$a=a_{v}+\delta a>a_{v}$ for a
small enough $\delta a$.
\subsection{\label{s:pt}Phase transitions and the effective vacuum density}
Since a substantial density of matter begins to arise only at $a=a_{0}$,
the corresponding moment of time could thus be associated with what is 
commonly referred to as the Big Bang. Although it could formally take an 
infinite time to reach $a_{0}$ from an infinitesimal $a$, such a reference
may be justified as marking the stage of a major transition of $\rho_{v}^{0}$
to matter. [As mentioned in Sec. \ref{s:mo} above, Eqs. (\ref{eq:f1}) and 
(\ref{eq:f2}) are invariant with respect to a shift in time. It still seems 
to be most consistent with the common perception to associate both the Big 
Bang and the zero time with the beginning of a major transition of 
vacuum to matter \cite{r:vd,r:bp}, which is identified by
manifestation of the nonlinearity.] 

Let $\rho_{r}(a)$ and $\rho_{n}(a)$ be the respective energy
densities of radiation and nonrelativistic matter. Let us also allow
for the equations of state for radiation, $p_{r}(a)=(1/3)\rho_{r}(a)$,
and for nonrelativistic matter, $p_{n}(a)=0$, where $p_{r}(a)$ and $p_{n}(a)$
are the pressures of radiation and nonrelativistic matter, respectively. 
One can then use additivity of the overall density and pressure as well: 
$\rho(a)=\rho_{v}(a)+\rho_{m}(a)$ and $p(a)=p_{v}(a)+p_{m}(a)$, where 
$p_{v}(a)=-\rho_{v}(a)$, $\rho_{m}(a)=\rho_{r}(a)+\rho_{n}(a)$, and
$p_{m}(a)=p_{r}(a)+p_{n}(a)$[ $=p_{r}(a)$]. Let the value of $a$ at
which vacuum decay (and the expansion acceleration) has to end be 
also denoted as $a_{m}\leq\infty$. As discussed in Sec. \ref{s:gc}
above, therefore, $\rho_{v}(a_{m})=0$. With Eq. (\ref{eq:den0}), 
integration of Eq. (\ref{eq:f3}) from $a=0$
to $a=a_{m}$ thus leads to:
\begin{equation}
\rho_{v}^{0}=\rho_{vr}^{0}+\rho_{vn}^{0}=
4\int_{0}^{a_{m}}\frac{\rho_{r}(a)}{a}da+
3\int_{0}^{a_{m}}\frac{\rho_{n}(a)}{a}da +\rho_{m}(a_{m}).
\label{eq:deninf}
\end{equation}

For $a_{1}<a_{m}$, however, Eq. (\ref{eq:deninf}) could be rewritten as
\begin{displaymath}
\rho_{v}^{0}=\int_{0}^{a_{1}}\frac{4\rho_{r}(a)+3\rho_{n}(a)}{a}da+
\int_{a_{1}}^{a_{m}}\frac{4\rho_{r}(a)+3\rho_{n}(a)}{a}da+\rho_{m}(a_{m})=
\end{displaymath}
\begin{equation}
\rho_{v}^{0}-\rho_{v}(a_{1})-\rho_{m}(a_{1})+\rho_{m}(a_{m})+
\int_{a_{1}}^{a_{m}}\frac{4\rho_{r}(a)+3\rho_{n}(a)}{a}da,
\label{eq:denp}
\end{equation}
where integration of Eq. (\ref{eq:f3}) from $0$ to $a_{1}$ has been used 
to calculate the first integral in Eq. (\ref{eq:denp}). At $a=a_{1}$,
the effective vacuum density $\rho_{v}^{e}(a_{1})=\rho_{v}(a_{1})$
could thus be written as
\begin{equation}
\rho_{v}^{e}(a_{1})=
\int_{a_{1}}^{a_{m}}\frac{4\rho_{r}(a)+3\rho_{n}(a)}{a}da+
\rho_{m}(a_{m})-\rho_{m}(a_{1}).
\label{eq:dene}
\end{equation}

Since $\rho_{v}^{e}(a_{1})$ has to be positive, the two first (definitely
positive) terms in the right hand side of Eq. (\ref{eq:dene}) have to exceed
$\rho_{m}(a_{1})$. An additional implication of $\rho_{v}^{e}(a_{1})>0$ for the
properties of vacuum decay could be seen from considering expansion of only such
matter amount as is available at $a=a_{1}$ (i.e., without an additional vacuum decay
into matter for $a>a_{1}$), $\rho_{m}^{1}(a)\equiv\rho_{r}^{1}(a)+\rho_{n}^{1}(a)$,
where $\rho_{r}^{1}(a)\equiv\rho_{r}(a_{1})a_{1}^{4}/a^{4}$ is the respective 
radiation density, and $\rho_{n}^{1}(a)\equiv\rho_{n}(a_{1})a_{1}^{3}/a^{3}$
is such density of nonrelativistic matter. $\rho_{v}^{e}(a_{1})>0$
thus implies that
\begin{displaymath}
\rho_{v}^{e}(a_{1})=
\int_{a_{1}}^{a_{m}}\frac{4[\rho_{r}(a)-\rho_{r}^{1}(a)]+
3[\rho_{n}(a)-\rho_{n}^{1}(a)]}{a}da+
\int_{a_{1}}^{a_{m}}\frac{4\rho_{r}^{1}(a)+3\rho_{n}^{1}(a)}{a}da+
\rho_{m}(a_{m})-\rho_{m}(a_{1})
\end{displaymath}
\begin{equation}
=\int_{a_{1}}^{a_{m}}\frac{\rho_{r}(a)-\rho_{r}^{1}(a)}{a}da+
3\int_{a_{1}}^{a_{m}}\frac{\rho_{m}(a)-\rho_{m}^{1}(a)}{a}da+
\rho_{m}(a_{m})-\rho_{m}^{1}(a_{m})>0.
\label{eq:dened}
\end{equation}
Since $\rho_{m}(a)>\rho_{m}^{1}(a)$ for $a>a_{1}$, only the first (integral) difference 
term in the last expression for the left hand side of inequality (\ref{eq:dened})
may not be positive. This could happen if transition from radiation to nonrelativistic
matter takes place and when such a transition is overall more effective than vacuum
decay into radiation for $a\in[a_{1},a_{m}]$. Even then, however, the other
two (definitely positive) difference terms have to outweigh 
the first one to ensure $\rho_{v}^{e}(a_{1})>0$.

When no transition from radiation to nonrelativistic matter takes 
place, Eq. (\ref{eq:dened}) for $\rho_{v}^{e}(a_{1})$ thus exposes 
the right hand side of Eq. (\ref{eq:dene}) as being generated by a
decay of the residual vacuum density between $a=a_{1}$ and $a=a_{m}$.
This vacuum density has been left after the preceding stages of phase
transitions. These stages are characterized by transition from the 
initial vacuum phase of the whole matter [Eq. (\ref{eq:den0})] to
the stage with the overall amount of both relativistic and 
nonrelativistic matter components at $a=a_{1}$
apart from $\rho_{v}^{e}(a_{1})$.

The expansion would be decelerating (accelerating) \cite{r:rp,r:prs,r:trgl} at 
$a=a_{1}$ if $2\rho_{v}^{e}(a_{1})<$ ( $>$ ) $\rho_{m}(a_{1})+\rho_{r}(a_{1})$
[$=\rho_{m}(a_{1})+3p_{m}(a_{1})$]. In view of Eqs. (\ref{eq:dene}) and
(\ref{eq:dened}), this is equivalent to
\begin{equation}
\int_{a_{1}}^{a_{m}}\frac{4\rho_{r}(a)+3\rho_{n}(a)}{a}da+
\rho_{m}(a_{m})<\textstyle{(}>\textstyle{)}\displaystyle\,\,\frac{3\rho_{m}(a_{1})+\rho_{r}(a_{1})}{2}\Longleftrightarrow
\label{eq:ae}
\end{equation}
\begin{equation}
\int_{a_{1}}^{a_{m}}\frac{[\rho_{r}(a)-\rho_{r}^{1}(a)]+
3[\rho_{m}(a)-\rho_{m}^{1}(a)]}{a}da+\rho_{m}(a_{m})-\rho_{m}^{1}(a_{m})<
\textstyle{(}>\textstyle{)}\displaystyle\, \frac{\rho_{m}(a_{1})+\rho_{r}(a_{1})}{2}.
\label{eq:aed}
\end{equation}
Since matter creation has to be most intensive during the initial manifestation of the
nonlinearity (when transition from radiation to nonrelativistic matter might also be 
taking place), the stage of nonlinear deceleration could be expected to eventually
switch to that of such an acceleration. This corresponds to $a_{1}$ moving from the
initial to later nonlinear stages of universe evolution up until $a_{1}=a_{m}$, where
$\rho_{m}(a_{m})+\rho_{r}(a_{m})>0$ for $a_{m}<\infty$ (if such a finite $a_{m}$ exists)
implies that a deceleration stage returns. The same order for $2\rho_{v}^{e}(a_{1})$
and $\rho_{m}(a_{1})+\rho_{r}(a_{1})$ [and then also for $\rho_{v}^{e}(a_{1})$ and
$\rho_{m}(a_{1})$] could thus be anticipated when both these magnitudes decrease with
$a_{1}$ increasing from such value of the scale factor as has made them equal. When 
$a_{1}$ is the value of $a$ relevant to the observations \cite{r:rp,r:prs,r:trgl}, 
the observed relation between $\rho_{v}^{e}(a_{1})$ and $\rho_{m}(a_{1})$ then 
provides another constraint for vacuum decay in terms of functions $\rho_{r}(a)$
and $\rho_{n}(a)$ for $a\in[a_{1},a_{m}]$ along with the ratio $(a_{m}/a_{1})$
in Eqs. (\ref{eq:dene}) and (\ref{eq:dened}). 
\section{\label{s:cc}The \mbox{Einstein} equations}
The \mbox{Einstein} equations \cite{r:ll} with the effective cosmological term
specified for the FRW metric by either of Eqs. (\ref{eq:dene}) and (\ref{eq:dened})
could thus be written as [signature $(+,-,-,-)$ is assumed]
\begin{equation}
R_{ij}-\frac{1}{2}g_{ij}R=8\pi G T_{ij},
\label{eq:en}
\end{equation}
where $g_{ij}$, $R_{ij}$, and $T_{ij}$ ($T_{i;j}^{j}=0$) are the metric,
\mbox{Ricci}, and total matter energy-momentum tensors, respectively,
and $R$ is the \mbox{Ricci} scalar curvature. When all phases 
of matter by definition described with $T_{ij}$ are present,
\begin{equation}
T_{ij}=\tilde{T}_{ij}+g_{ij}\rho_{v}^{e}=g_{ij}(\rho_{v}^{e}-p_{m})+
(\rho_{m}+p_{m})u_{i}u_{j},
\label{eq:emc}
\end{equation}
where $\rho_{v}^{e}$ is the effective energy density of vacuum defined for
the FRW $a=a_{1}$ by either of Eqs. (\ref{eq:dene}) and (\ref{eq:dened}),
and 
\begin{equation}
\tilde{T}_{ij}=(\rho_{m}+p_{m})u_{i}u_{j}-g_{ij}p_{m}
\label{eq:emm}
\end{equation}
is the energy-momentum tensor of the matter phase. In Eqs. (\ref{eq:emc})
and (\ref{eq:emm}), $\rho_{m}=\rho_{r}+\rho_{n}$ and $p_{m}=p_{r}+p_{n}$ 
[$p_{r}=(1/3)\rho_{r}$ and $p_{n}=0$] are the energy density and pressure,
respectively, of the matter phase, and $u_{i}$ is the four-velocity.
Before the initial stage of phase transitions, only vacuum is
present. In view of Eq. (\ref{eq:den0}), Eq. (\ref{eq:emc})
then has to look as 
\begin{equation}
T_{ij}=g_{ij}\rho_{v}^{0}=g_{ij}(\rho_{vr}^{0}+\rho_{vn}^{0}).
\label{eq:emi}
\end{equation}

If the right hand side of Eq. (\ref{eq:den0})
also contained a part of the vacuum energy density that is not the origin of 
any matter, say $\rho_{v}^{v}>0$, this vacuum density could not then stand for a
phase of matter. Since $\rho_{v}^{v}$ could thus be viewed as a constant of nature, 
Eqs. (\ref{eq:emc}) and (\ref{eq:emi}) legitimately imply that the corresponding term,
$g_{ij}\lambda$ with $\lambda\equiv 8\pi G\rho_{v}^{v}$, would then have to be 
present in the left hand side of Eqs. (\ref{eq:en}) as well. The possibility 
of having such a $\lambda$-term in the left hand side of Eqs. (\ref{eq:en})
is thus reserved for offsetting the effect of $\rho_{v}^{v}$
in the right hand side of these equations.

Such a formulation of Eqs. (\ref{eq:en}) implies that a vacuum density
matters gravitationally only when it arises from matter in the framework of these
equations. The absence of $\rho_{v}^{v}$ in Eqs. (\ref{eq:emc}) and (\ref{eq:emi})
is then due to the failure of vacuum with $\rho_{v}^{v}$ to be the origin of
matter. As discussed in Sec. {\ref{s:pi} above, such a vacuum could only correspond
to the \mbox{Friedmann} scale factor $a=0$. This would be inconsistent with the growth 
of $a>0$ actually characterizing the expanding universe. $\rho_{v}^{v}$ could not then
be relevant to such a growth, for its space-time response also has to be constant when
$a$ is viewed as a measure of the phase transition as well. A $\lambda$-term,
$g_{ij}\lambda$, in the left hand side of Eqs. (\ref{eq:en}) would also have
to be only a(n equal) space-time response to the presence of $8\pi G g_{ij}\rho_{v}^{v}$
in the right hand side of these equations. The term $8\pi G g_{ij}\rho_{v}^{v}$ in
Eqs. (\ref{eq:en}) thus has to either be zero or be canceled by its simultaneous
presence in both sides of these equations. 
\section{Conclusions}
It follows from the \mbox{Friedmann} equations that a universe being capable
of avoiding an infinite initial energy density could only be spatially flat. 
Such equations with a finite initial density then have to be valid when the scale
factor is equal to zero as well. They thus rigorously require that the corresponding
universe be in the state of vacuum when the scale factor is infinitesimal. Matter
creation is therefore largely the result of such a finite increase of the scale
factor as necessitates the manifestation of respective nonlinear effects 
in the \mbox{Friedmann} equations. The beginning of the nonlinear
stage of the matter-creating phase transitions could also be
referred to as the Big Bang.

Correspondence between the homogeneous density of a matterless vacuum 
and a matter-density distribution in the respective \mbox{Friedmann}
universe whose zero-$a$ limit is represented by such a vacuum is thus 
suggested by the \mbox{Einstein} equations. In the most general form,
this correspondence implies that a vacuum density has to be gravitationally
significant in these equations only when it could have a respective matter
phase as well. It thus also means that vacuum decay into matter has to be 
continuing at the stage of accelerated expansion. The effective cosmological
term is then specified by the vacuum phase of matter that has been left 
after the preceding phase transitions from vacuum to matter.


\begin{thebibliography}{00}
\bibitem{r:rp}{S.~Weinberg,}
{Rev. Mod. Phys. {\bf 61}, 1--22 (1989);\newline 
S.~M.~Carroll, W.~H.~Press, E.~L.~Turner,
Annu. Rev. Astron. Astrophys. {\bf 30}, 499--542 (1992);\newline
V.~Sahni, A.~Starobinsky,
Int. J. Mod. Phys. D {\bf 9}, 373--443 (2000);\newline
S.~M.~Carroll,
Liv. Rev. Rel. {\bf 4}, 1--56 (2001);\newline 
P.~J.~E.~Peebles, B.~Ratra,
Rev. Mod. Phys. {\bf 75}, 559--606 (2003);\newline
T.~Padmanabhan, 
Phys. Rep. {\bf 380}, 235--320 (2003);\newline
S.~Nobbenhuis,
Found. Phys. {\bf 36}, 613--680 (2006);\newline
J.~A.~Frieman, M.~S.~Turner, D.~Huterer,
Annu. Rev. Astron. Astrophys. {\bf 46}, 385--432 (2008);\newline
R.~Bousso,
Gen. Rel. Grav. {\bf 40}, 607--637 (2008);\newline
T.~Padmanabhan,
Gen. Rel. Grav. {\bf 40}, 529--564 (2008);\newline
A.~D.~Chernin,
Phys. Usp. {\bf 51}, 253--282 (2008);\newline
R.~R.~Caldwell, M.~Kamionkowski,
Annu. Rev. Nucl. Part. Sci. {\bf 59}, 397--429 (2009);\newline
M.~Li, X.~D.~Li, S.~Wang, Y.~Wang,
Comm. Theor. Phys. {\bf 56}, 525--604 (2011);\newline
Yu.~L.~Bolotin, D.~A.~Erokhin, O.~A.~Lemets,
Phys. Usp. {\bf 55}, 876--918 (2012);\newline
as with the other references below, the above list of reviews is hoped to be fairly
representative whereas it is far from being complete.}
\bibitem{r:prs}{S.~Perlmutter {\em et al.} (Supernova Cosmology Project Collaboration),}
{Astrophys. J. {\bf 517}, 565--586 (1999);\newline
A.~G.~Riess {\em et al.} (Supernova Search Team Collaboration),
Astrophys. J. {\bf 116}, 1009--1038 (1998).}
\bibitem{r:trgl}{N.~A.~ Bahcall, J.~P.~Ostriker, S.~Perlmutter, P.~J.~Steinhardt,}
{Science {\bf 284}, 1481--1488 (1999);\newline
W.~L.~Freedman, M.~S.~Turner,
Rev. Mod. Phys. {\bf 75}, 1433--1447 (2003);\newline
M.~Bartelmann, 
Rev. Mod. Phys. {\bf 82}, 331--382 (2010);\newline
N.~Jarosik {\em et al.},
Astrophys. J. Suppl. Ser. {\bf 192}, 14 (2011);\newline
E.~Komatsu {\em et al.},
Astrophys. J. Suppl. Ser. {\bf 192}, 18 (2011).}
\bibitem{r:g}{E.~B.~Gliner,}
{Sov. Phys. JETP {\bf 22}, 378--382 (1966).}
\bibitem{r:s}{A.~D.~Sakharov,}
{Sov. Phys. JETP {\bf 22}, 241--249 (1966).}
\bibitem{r:z}{Y.~B.~Zeldovich,}
{JETP Lett. {\bf 6}, 316--317 (1967);\newline
Y.~B.~Zeldovich,
Sov. Phys. Usp. {\bf 11}, 381--393 (1968);\newline
Y.~B.~Zeldovich,
Sov. Phys. Usp. {\bf 24}, 216--230 (1981).}
\newpage
\bibitem{r:tdm}{J.~D.~Bekenstein,}
{Phys. Rev. D {\bf 7}, 2333--2346 (1973);\newline
J.~M.~Bardeen, B.~Carter, S.~W.~Hawking,
Comm. Math. Phys. {\bf 31}, 161--170 (1973);\newline
S.~W.~Hawking,
Comm. Math. Phys. {\bf 43}, 199--220 (1975);\newline
G.~W.~Gibbons, S.~W.~Hawking,
Phys. Rev. D {\bf 15}, 2738--2751 (1977);\newline
G.~t'Hooft,
arXiv:gr-qc/9310026;\newline
L.~Susskind,
J. Math Phys. {\bf 36}, 6377--6396 (1995);\newline
T.~Jacobson,
Phys. Rev. Lett. {\bf 75}, 1260--1263 (1995);\newline
R.~M.~Wald,
Liv. Rev. Rel. {\bf 4}, 1--44 (2001);\newline
R.~Bousso,
Rev. Mod. Phys. {\bf 74}, 825--874 (2002);\newline
T.~Padmanabhan,
Phys. Rep. {\bf 406}, 49--125 (2005);\newline
T.~Padmanabhan,
Rep. Prog. Phys. {\bf 73}, 046901 (2010);\newline
E.~Verlinde,
J. High Energy Phys. 04 (2011) 029;\newline
D.~A.~Easson, P.~H.~Frampton, G.~F.~Smoot,
Phys. Lett. B {\bf 696}, 273--277 (2011);\newline
T.~Padmanabhan,
Res. Astron. Astrophys. {\bf 12}, 891--916 (2012);\newline
see also a related discussion in the last paper of Ref. \cite{r:rp};\newline
N.~Komatsu, S.~Kimura,
Phys. Rev. D {\bf 87}, 043531 (2013) and recent references therein.}
\bibitem{r:vd}{K.~Freese, F.~C.~Adams, J.~A.~Frieman, E.~Mottola,}
Nucl. Phys. B {\bf 287}, 797--814 (1987);\newline
P.~J.~E.~Peebles, B.~Ratra,
Astrophys. J. {\bf 325}, L17--L20 (1988);\newline
J.~A.~S.~Lima,
Phys. Rev. D {\bf 54}, 2571--2577 (1996);\newline
J.~M.~Overduin, F.~I.~Copperstock,
{Phys. Rev. D {\bf 58}, 043506 (1998);\newline
J.~M.~Overduin,
Astrophys. J. {\bf 517}, L1--L4 (1999);\newline
H.~A.~Borges, S.~Carneiro,
Gen. Rel. Grav. {\bf 37}, 1385--1394 (2005);\newline
J.~S.~Alcaniz, J.~A.~S.~Lima,
Phys. Rev. D {\bf 72}, 063516 (2005);\newline
J.~D.~Barrow, T.~Clifton, 
Phys. Rev. D {\bf 73}, 103520 (2006);\newline
S.~Carneiro,
J. Phys. A {\bf 40}, 6841--6848 (2007);\newline
S.~Carneiro, M.~A.~Dantas, C.~Pigozzo, J.~S.~Alcaniz,
Phys. Rev. D {\bf 77}, 083504 (2008);\newline
J.~S.~Alcaniz, H.~A.~Borges, S.~Carneiro, J.~C.~Fabris, C.~Pigozzo, W.~Zimdahl,
Phys. Lett. B {\bf 716}, 165--170 (2012);\newline
J.~A.~S.~Lima, S.~Basilakos, J.~Sola,
Mon. Not. R. Astron. Soc. {\bf 431}, 923--929 (2013);\newline
see also the relevant references in these papers as well as
in the last paper of Ref. \cite{r:tdm}.}
\bibitem{r:zmb}{Y.~B.~Zeldovich,}
{JETP Lett. {\bf 12}, 307--311 (1970);\newline
S.~Weinberg,
Astrophys. J. {\bf 168}, 175--194 (1971);\newline
G.~L.~Murphy,
Phys. Rev. D {\bf 8}, 4231--4233 (1973);\newline
J.~D.~Barrow,
Phys. Lett. B {\bf 180}, 335--339 (1986);\newline
T.~Padmanabhan, S.~M.~Chitre,
Phys. Lett. A {\bf 120}, 433--436 (1987);\newline
W.~Zimdahl,
Phys. Rev. D {\bf 53}, 5483--5493 (1996);\newline
W.~Zimdahl, D.~J.~Schwarz, A.~B.~Balakin, D.~Pavon,
Phys. Rev. D. {\bf 64}, 063501 (2001);\newline
I.~Brevik, O.~Gorbunova,
Gen. Rel. Grav. {\bf 37}, 2039--2045 (2005);\newline
J.~C.~Fabris, S.~V.~B.~Goncalves, R.~D.~Ribeiro,
Gen. Rel. Grav. {\bf 38}, 495--506 (2006);\newline
R.~Colistete, J.~C.~Fabris, J.~Tossa, W.~Zimdahl,
Phys. Rev. D {\bf 76}, 103516 (2007);\newline
B.~Li, J.~D.~Barrow,
Phys. Rev. D {\bf 79}, 103521 (2009);\newline
W.~S.~Hip\'{o}lito-Ricaldi, H.~E.~S.~Velten, W.~Zimdahl,
Phys. Rev. D {\bf 82}, 063507 (2010);\newline
J.-S.~Gagnon, J.~Lesgourgues,
J. Cosmol. Astropart. Phys. 09 (2011) 026;\newline
see also the relevant references in these papers as 
well as in the last paper of Ref. \cite{r:tdm}.}
\newpage
\bibitem{r:pn}{I.~Prigogine, J.~Geheniau, E.~Gunzig, P.~Nardone,}
{Gen. Rel. Grav. {\bf 21}, 767--776 (1989);\newline
M.~O.~Calv\~{a}o, J.~A.~S.~Lima, I.~Waga,
Phys. Lett. A {\bf 162}, 223--226 (1992);\newline
J.~A.~S.~Lima, A.~S.~M.~Germano,
Phys. Lett. A {\bf 170}, 373--378 (1992);\newline
J.~A.~S.~Lima, A.~S.~M.~Germano, L.~R.~W.~Abramo,
Phys. Rev. D {\bf 53}, 4287--4297 (1996);\newline
J.~A.~S.~Lima, L.~R.~W.~Abramo,
Phys. Lett. A {\bf 257}, 123--131 (1999);\newline
J.~A.~S.~Lima, F.~E.~Silva, R.~C.~Santos, 
Class. Quant. Grav. {\bf 25}, 205006 (2008);\newline
G.~Steigman, R.~C.~Santos, J.~A.~S.~Lima,
J. Cosmol. Astropart. Phys. 06 (2009) 033;\newline
J.~A.~S.~Lima, J.~F.~Jesus, F.~A.~Oliveira, 
J. Cosmol. Astropart. Phys. 11 (2010) 027;\newline
J.~A.~S.~Lima, S.~Basilakos, F.~E.~M.~Costa,
Phys. Rev. D {\bf 86}, 103534 (2012);\newline
see also the relevant references in these papers.}
\bibitem{r:gd}{E.~B.~Gliner,}
{Sov. Phys. Dokl. {\bf 15}, 559--562 (1970).}
\bibitem{r:bp}{H.~J.~Blome, M.~Priester,} 
{Astron. Astrophys. {\bf 250}, 43--49 (1991) and references therein.}
\bibitem{r:ir}{M.~Israelit, N.~Rosen,}
{Astrophys. Sp. Sci. {\bf 204}, 317--327 (1993).}
\bibitem{r:ll}{L.~D.~Landau, E.~M.~Lifshitz,}
{The classical theory of fields, Fizmatlit, Moscow (2003) (in Russian).}
\bibitem{r:kt}{E.~W.~Kolb, M.~S.~Turner,}
{The early universe. Addison-Wesley (1990).}
\bibitem{r:fn}{A.~Friedmann,}
{Z. Phys. {\bf 10}, 377--386 (1922) [Gen. Rel. Grav. {\bf 31}, 1991--2000 (1999)];\newline
A.~Friedmann, 
Z. Phys. {\bf 21}, 326--332 (1924) [Gen. Rel. Grav. {\bf 31}, 2001--2008 (1999)].}
\bibitem{r:mr}{M.~A.~Markov,}
{JETP Lett. {\bf 36}, 265--267 (1982);\newline
N.~Rosen,
Astrophys. J. {\bf 297}, 347--349 (1985).}
\bibitem{r:hpg}{R.~Penrose,}
{Phys. Rev. Lett. {\bf 14}, 57--59 (1965);\newline
S.~W.~Hawking, 
Phys. Rev. Lett. {\bf 15}, 689--690 (1965);\newline
S.~W.~Hawking,
Phys. Rev. Lett. {\bf 17}, 444--445 (1966);\newline
R.~P.~Geroch,
Phys. Rev. Lett. {\bf 17}, 445--447 (1966);\newline
S.~W.~Hawking,
Proc. R. Soc. Lond. Ser. A {\bf 294}, 511--521 (1966);\newline
S.~W.~Hawking,
Proc. R. Soc. Lond. Ser. A {\bf 295}, 490--493 (1966);\newline
S.~W.~Hawking,
Proc. R. Soc. Lond. Ser. A {\bf 300}, 187--201 (1967);\newline
R.~Geroch,
Ann. Phys. {\bf 48}, 526--540 (1968);\newline
S.~W.~Hawking, G.~F.~R.~Ellis,
Astrophys. J. {\bf 152}, 25--36 (1968);\newline
S.~W.~Hawking, R.~Penrose,
Proc. R. Soc. Lond. Ser. A {\bf 314}, 529--548 (1970).}
\bibitem{r:dm}{E.~Poisson, W.~Israel,}
{Class. Quant. Grav. {\bf 5}, L201--L205 (1988);\newline
I.~Dymnikova,
Gen. Rel. Grav. {\bf 24}, 235--242 (1992);\newline
P.~O.~Mazur, E.~Mottola,
Proc. Nat. Acad. Sci. USA {\bf 101}, 9545--9550 (2004);\newline
I.~Dymnikova, E.~Galaktionov,
Class. Quant. Grav. {\bf 22}, 2331--2357 (2005) and references therein.}
\bibitem{r:t}{E.~P.~Tryon,}
{Nature {\bf 246}, 396--397 (1973).}
\bibitem{r:psz}{L.~Parker,}
{Phys. Rev. Lett. {\bf 21}, 562--564 (1968);\newline
L.~Parker,
Phys. Rev. {\bf 183}, 1057--1068 (1969);\newline
R.~U.~Sexl, H.~K.~Urbantke,
Phys. Rev. {\bf 179}, 1247--1250 (1969);\newline
the first paper in Ref. \cite{r:zmb};\newline
L.~Parker,
Phys. Rev. {\bf 3}, 346--356 (1971); Phys. Rev. {\bf 3}, 2546--2546 (1971);\newline
Y.~B.~Zeldovich, L.~P.~Pitaevskii,
Comm. Math. Phys. {\bf 23}, 185--188 (1971);\newline
the fourth paper (by G.~W.~Gibbons and S.~W.~Hawking) in Ref. \cite{r:tdm};\newline
L.~Parker,
J. Phys. A {\bf 45}, 374023 (2012) and references therein.}
\bibitem{r:skhp}{A.~D.~Sakharov,}
{Sov. Phys. JETP {\bf 52}, 349--351 (1980);\newline
R.~Peierls, Phys. Today {\bf 47}(11), 115--115 (1994).}
\bibitem{r:aqn}{H.~Alfv\'{e}n,}
{Rev. Mod. Phys. {\bf 37}, 652--665 (1965);\newline
H.~Alfv\'{e}n,
Phys. Today {\bf 24}(2), 28--33 (1971);\newline
G.~Steigman,
Annu. Rev. Astron. Astrophys. {\bf 14}, 339--372 (1976);\newline
F.~Wilczek,
Sci. Am. {\bf 243}(6), 82--90 (1980);\newline
H.~R.~Quinn,
Phys. Today {\bf 56}(2), 30--35 (2003);\newline
M.~Dine, A.~Kusenko,
Rev. Mod. Phys. {\bf 76}, 1--30 (2004);\newline
G.~Steigman,
J. Cosmol. Astropart. Phys. 10 (2008) 001;\newline
L.~Canetti, M.~Drewes, M.~Shaposhnikov,
New J. Phys. {\bf 14}, 095012 (2012).}
\bibitem{r:grn}{O.~Gron,}
{Am. J. Phys. {\bf 54}, 46--52 (1986).}
\bibitem{r:inf}{A.~H.~Guth,}
{Phys. Rev. D {\bf 23}, 347--356 (1981);\newline
K.~Sato,
Mon. Not. R. Astron. Soc. {\bf 195}, 467--479 (1981);\newline
A.~D.~Linde,
Rep. Prog. Phys. {\bf 47}, 925--986 (1984);\newline
J. V. Narlikar, T. Padmanabhan,
Annu. Rev. Astron. Astrophys. {\bf 29}, 325--362 (1991);\newline 
A.~Linde,
Lect. Not. Phys. {\bf 738}, 1--54 (2008).}
\bibitem{r:infdg}{I.~G.~Dymnikova,}
{Sov. Phys. JETP {\bf 63}, 1111--1115 (1986);\newline
E.~B.~Gliner,
Phys. Usp. {\bf 45}, 213--220 (2002).}
\bibitem{r:infps}{R.~Penrose,}
{Ann. N. Y. Acad. Sci. {\bf 571}, 249--264 (1989);\newline
P.~J.~Steinhardt,
Sci. Am. {\bf 304}(4), 36--43 (2011).}
\bibitem{r:grn02}{O.~Gron,}
{Eur. J. Phys. {\bf 23}, 135--144 (2002).}
\bibitem{r:ls}{J.~A.~S.~Lima, J.~Santos,}
{Int. J. Theor. Phys. {\bf 34}, 127--134 (1995).}
\bibitem{r:vdo}{C.~Shapiro, M.~S.~Turner,}
{Astrophys. J. {\bf 649}, 563--569 (2006);\newline
A.~Shafieloo, V.~Sahni, A.~A.~Starobinsky,
Annalen der Physik {\bf 19}, 316--319 (2010);\newline
A.~C.~C.~Guimar\~{a}es, J.~A.~C.~Lima,
Class. Quant. Grav. {\bf 28}, 125026 (2011).}
\end{thebibliography}
\end{document}